\documentstyle[11pt]{article}                   % [bkp.tex 28.05.1997]
\def\quality{\textheight=240mm \textwidth=160mm \hoffset=-20mm 
             \topmargin 0Truein} %[11pt]
\quality

\def\IR{\hbox{\rm I\kern-.2em\hbox{\rm R}}}
\def\IL{\hbox{\rm I\kern-.2em\hbox{\rm L}}}
\def\IZ{\hbox{{\rm Z}\kern-.3em{\rm Z}}}      
\def\const{\, {\rm Const} \,}   \def\mod1{\,({\rm mod\ } 1)\,}
\def\ep{\varepsilon}        \def\phi{\varphi}        \def\al{\alpha}
\def\be{\begin{equation}}   \def\ee{\end{equation}}  
\def\beq#1#2{\begin{equation} \label{#1} #2 \end{equation}}
\def\bea#1{\begin{eqnarray*} #1 \end{eqnarray*}} \def\a{\!\!\!&\!\!\!\!&}
\def\function#1{\left\{\!\!\!\begin{array}{ll} #1 \end{array} \right.}
\def\lrp#1{\left( #1 \right)} % (#1)

\newtheorem{theorem}{Theorem}[section]   %Numbering: Theorem--Other section
\newtheorem{lemma}{Lemma}[section]       %{lemma}[theorem]{Lemma}   section
\newtheorem{prop}{Proposition}[section]
\newtheorem{dftn}{Definition}[section]
\newenvironment{definition}{\begin{dftn}\rm}{\end{dftn}} %section
             \catcode`@=11 \@addtoreset{equation}{section} \catcode`@=12
\def\proof{\smallskip \noindent {\bf Proof. \ }}
\newcommand\filledsquare{\ \vrule width 1.5ex height 1.2ex}  %filled square
\def\qed{\hfill\filledsquare\linebreak\smallskip\par}
\makeatletter \newcounter{@sc} \newcounter{@scp} \newcounter{@t} \newlength{\@x}
\newlength{\@xa} \newlength{\@xb}  \newlength{\@y} \newlength{\@ya}
\newlength{\@yb} \newsavebox{\@pt}
\def\bezier#1(#2,#3)(#4,#5)(#6,#7){\c@@sc#1\relax
 \c@@scp\c@@sc \advance\c@@scp\@ne
 \@xb #4\unitlength \advance\@xb -#2\unitlength \multiply\@xb \tw@
 \@xa #6\unitlength \advance\@xa -#2\unitlength
 \advance\@xa -\@xb \divide\@xa\c@@sc
 \@yb #5\unitlength \advance\@yb -#3\unitlength \multiply\@yb \tw@
 \@ya #7\unitlength \advance\@ya -#3\unitlength
 \advance\@ya -\@yb \divide\@ya\c@@sc
 \setbox\@pt\hbox{\vrule height\@halfwidth depth\@halfwidth
 width\@wholewidth}\c@@t\z@
 \put(#2,#3){\@whilenum{\c@@t<\c@@scp}\do
 {\@x\c@@t\@xa \advance\@x\@xb \divide\@x\c@@sc \multiply\@x\c@@t
 \@y\c@@t\@ya \advance\@y\@yb \divide\@y\c@@sc \multiply\@y\c@@t
 \raise \@y \hbox to \z@{\hskip \@x\unhcopy\@pt\hss}\advance\c@@t\@ne}}}
\makeatother
\def\Bfig(#1,#2)#3#4{\begin{figure} \begin{center}
    \setlength{\unitlength}{\mlbscale} \begin{picture}(#1,#2) #3
    \end{picture} \end{center} \caption{#4} \end{figure}}
\newcommand\mlbscale{1pt}

\newif\iffigs \figsfalse 
\def\drawing #1 #2 #3 
   {\begin{center} \setlength{\unitlength}{1pt}
    \begin{picture}(#1,#2)(0,0) 
    \put(0,0){\framebox(#1,#2){#3}} \end{picture} \end{center}}
\begin{document}

\title{A KAM type theorem for systems with round-off errors}
\author{M. Blank\thanks{Russian Academy of Sciences, 
                Inst. for Information Transmission Problems,
                B.Karetnij Per. 19, 101447, Moscow, Russia.},
        T. Kr\"uger, L. Pustyl'nikov
        \\ Forschungszentrum BiBoS, Universitat Bielefeld \\ 
        D-33615 Bielefeld, Germany.}
\date{\today}
\maketitle

\noindent{\bf Abstract.} 
Perturbations due to round-off errors in computer modeling are
discontinuous and therefore one cannot use results like KAM theory
about smooth perturbations of twist maps. We elaborate a special
approximation scheme to construct two smooth periodic on the angle
perturbations of the twist map, bounding the discretized map from
above and from below. Using the well known Moser's theorem we prove
the existence of invariant curves for these smooth approximations. As
a result we are able to prove that any trajectory of the discretized
twist map is eventually periodic. We discuss also some questions,
concerning the application of the intersection property in Moser's
theorem and the generalization of our results for the twist map in
Lobachevski plane.

\section{Introduction}

In recent years an increasing interest has been addressed to
problems related to numerical simulations of complex dynamics. 
Classical mathematical results in this field provide the theoretical 
foundation for the numerical modeling of stable dynamical systems, 
while the analysis of mathematical problems arising due to round-off 
errors was started relatively recently. Probably the first theoretical 
study was published in 1984 in \cite{Bl1}. The results of this paper 
show that the influence of arbitrary small round-off errors can change 
the behavior of a complex (chaotic) system drastically, and that the 
resulting behavior depends very sensitively on the fine structure 
of the phase space discretization corresponding to these errors. It is 
enough to mention here the period multiplication phenomenon 
\cite{Bl1}, and the localization under the action of such perturbations 
\cite{Bl1}-\cite{Bl19}, when trajectories that should normally be dense 
in the phase space remain confined to a small region. Starting from 1984 
a large number of publications were devoted to the mathematical analysis 
of problems related to numerical simulations of complex dynamics 
(see, for example, \cite{Bl5}--\cite{Vl} and references therein).

One of the most complex and intriguing problem arising in this field 
is the analysis of the influence of round-off errors for systems whose 
trajectories are neither stable nor unstable, or such that they have 
local components of this type. A typical example here is a rotation 
map $F: \; (\phi,r) \to (\phi+\al,r)$ of the two-dimensional plane $\IR^2$ 
around the origin. The study of the discretized version of this map was 
started in \cite{Bl16}. If the angle $\al \in [0,1)$ is irrational then 
any trajectory of the map $F$ is dense on a circle. However the action of 
round-offs  disrupts these invariant circles, and a'priori it is not 
clear that there are no trajectories going to infinity. Moreover in 
\cite{Bl16} an example of the rotation-like map with this property was 
constructed. Starting from that time at least 10 papers were dedicated 
to the discussion of this problem, however it was not solved.

%%%%%%%%%%%%%%%%%%%%%%%%%%%%%%%%%%%%%%%%%%%%%%%%%%%%%%%%%%%%%%%%%%%%%%%%%
%% PS-figure for the invariant set of the pure rotation
\begin{figure} \iffigs
   \input{psfig}
   \centerline{\psfig{file=rot.ps,width=220pt}}
   \centerline{\begin{picture}(0,0) \put(-108,15.5){\line(1,0){214.5}} 
               \end{picture}}
   \else \drawing 150 10 {Discretized plane rotation} \fi
   \caption{Invariant set (black dots) of the 0.0035-discretized pure plane 
            rotation} \label{d:rot} \end{figure}
%%%%%%%%%%%%%%%%%%%%%%%%%%%%%%%%%%%%%%%%%%%%%%%%%%%%%%%%%%%%%%%%%%%%%%%%%

Numerical simulations (see Figure~\ref{d:rot}) show that the invariant set 
of the discretized rotation looks very much like invariant tori in the 
Kolmogorov - Arnold - Moser (KAM) theory. Therefore in \cite{Bl19} it was 
conjectured that in spite of the discontinuities of the perturbations 
due to the round-off errors some kind of KAM theory can be constructed for 
this problem, namely to prove that any trajectory of the discretized rotation 
is eventually periodic. In this paper we follow this idea and the main result 
is the following theorem.

\begin{definition} \label{i:twist} A map $F$ of the annulus
$$ {\cal A}(r_-,r_+) :=
   \{ (\phi,r) \in \IR^2: \; 0 \le r_- \le r \le r_+ \le \infty \} $$
into $\IR^2$ is said to be {\em twist\/} if in polar coordinates
$(0 \le \phi < 1, \; 0 \le r <\infty)$ it can be written as
\beq{twist}{
    \function{&\!\!\! \phi \to \phi + \Phi(r) \mod1 \\
              &\!\!\! r    \to r} }
where the function $\Phi(r)$ is $C^l$-smooth for a sufficiently large $l$ 
($l=6$ is enough) and satisfies the inequality 
$$ |d\Phi(r)/dr|>0 .$$
\end{definition}

Notice, that in principle, the constant $r_+$ might be equal to $\infty$, 
in which case the annulus ${\cal A}(0,\infty)$ coincides with the entire 
plane $\IR^2$.

\begin{theorem} \label{t:twist} For any constants 
$0<\hat r_-<r_-<r_+<\hat r_+<\infty$ there exists $\ep_0>0$ such that 
for each $0<\ep<\ep_0$ if $r_-<r<r_+$ then
$\hat r_- < (D_\ep \circ F)^n (\phi,r) < \hat r_+$ for any $n \in \IZ^+$
and any angles $\phi \in [0,1)$. Moreover, if we additionally assume 
that $|d\Phi(r)/dr|>\const/r$ for any $r>0$, then for each $\ep>0$ any 
trajectory of the $\ep$-discretized twist map~(\ref{twist}) is eventually 
periodic. \end{theorem}

Perturbations due to round-off errors in computer modeling are
discontinuous and therefore one cannot use results like KAM theory about
smooth perturbations of twist maps. We elaborate a special approximation
scheme to construct two smooth periodic on $\phi$ perturbations of the
twist map, bounding the discretized map from above and from below. Using
the well known Moser's theorem \cite{Mos} we prove the existence of
invariant curves for these smooth approximations. The existence of the
invariant curves yields by some special shadowing argument the boundedness
of the trajectories of the discretized twist map.

\section{Twist maps with $C^\infty$-smooth perturbations}

Consider a family of $C^{\infty}$-perturbed twist maps of the annulus
${\cal A}(r_-,r_+)$
\beq{twist-0}{
    \function{&\!\!\! \phi \to \phi + \Phi(r) + \ep h(\phi,r,\ep) \mod1 \\
              &\!\!\! r    \to r    + \ep g(\phi,r,\ep) } }
depending on the small parameter $\ep \ge 0$.

In this section we intensively use the following well known result due to
Moser about the existence of invariant curves for the system~(\ref{twist-0}).

\begin{theorem} \cite{Mos} \label{Moser}
Let us assume that
\begin{itemize}
  \item $|d\Phi(r)/dr| > 0$ for any $r_- \le r \le r_+$, $r_+-r_->1$.
  \item The functions $\Phi,h,g \in C^\infty$ with respect to all variables,
        and $h$ and $g$ are $1$-periodic with respect to the angle variable 
        $\phi$.
  \item For a sufficiently small $\ep_0$ for any $|\ep| \le \ep_0$ any
        closed curve around the origin intersects with its image under the
        action of the map~(\ref{twist-0}) (intersection property).
\end{itemize}
Then for any $0 \le \ep \le \ep_0$ in the annulus $A$ there exists an
(\ref{twist-0})-invariant closed curve. \end{theorem}

Once again the assumption about the $C^\infty$-smoothness can be weaken to
the $C^l$-smoothness for a sufficiently large $l$ ($l=6$ is enough).

The last of the above assumptions is called the {\em intersection property}.
Actually in the proof of this theorem a weaker condition is verified. Namely,
it is enough to check up this condition only for closed curves
$\{(\phi,r): \; r=r(\phi)\}$ satisfying the inequality
$|dr/d\phi| < \ep^{3/8}$ (i.e., curves $C^1$-close to circles). This follows
from the estimate of the smoothness of the invariant curve and the fact that
the intersection property is used in the proof only to check up that after
the change of variables any circle intersects with its image (see \cite{Mos}
p.2,15). Therefore we can use this weaker condition as a definition of the
intersection property.

Moreover it is straightforward to show that if any circle centered at the 
origin intersects with its image than any closed curve intersecting two 
concentric circles with the distance $2\ep$ between them also intersects with 
its image under the action of the above map. 

It seems that using this argument one can weaken the intersection property
even more to the property that any circle centered at the origin intersects
with its image. Moreover one might think that the intersection property for
the circles yields this property for an arbitrary curve close to a circle.
However, the following lemma shows that this is not true.

\begin{lemma} \label{l:int.prop} There exists a $C^\infty$ map $f$ of the 
annulus ${\cal A}$, for which the intersection property holds for any circle 
centered at the origin, but there is a closed curve arbitrary close to a 
circle which does not intersect with its image. \end{lemma}

%%%%%%%%%%%%%%%%%%%%%%%%%%%%%%%%%%%%%%%%%%%%%%%%%%%%%%%%%%%%%%%%%%
%% Picture for the intersection property
\def\bline(#1,#2)(#3,#4)(#5){\put(#1,#2){\line(#3,#4){#5}}}
\Bfig(150,150){\bline(0,0)(1,0)(150)   \bline(0,0)(0,1)(150)
               \bline(0,150)(1,0)(150) \bline(150,0)(0,1)(150)
               \bezier{221}(20,85)(75,180)(130,85)
               \bezier{201}(20,65)(75,-30)(130,65)
               \bezier{53}(20,85)(10,75)(20,65)
               \bezier{53}(130,85)(140,75)(130,65) %end or circle
               \bezier{221}(65,35)(-25,75)(65,115)
               \bezier{221}(85,35)(175,75)(85,115)
               \bezier{53}(65,35)(75,25)(85,35)
               \bezier{53}(65,115)(75,125)(85,115)
               \put(110,20){$\Gamma$} \put(75,50){$f(\Gamma)$}
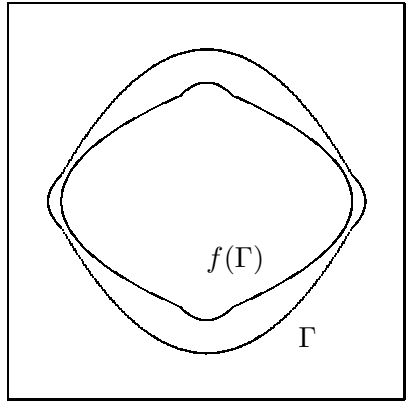
              }{Counterexample for the intersection property.
                \label{mos.int}}
%%%%%%%%%%%%%%%%%%%%%%%%%%%%%%%%%%%%%%%%%%%%%%%%%%%%%%%%%%%%%%%%%%

\proof Let the map $f$ be a superposition of the rotation by $\pi/2$ around
the origin and the linear map $S$, contracting with the coefficient $1-a$ on
the vertical coordinate ($0<a\ll1$). Notice that we first apply the rotation.
Clearly under the action of this map any circle centered at the origin
intersects with its image at points with angles $0$ and $\pi$. Consider now
a curve $\Gamma$, coinciding with a circle everywhere except small
neighborhoods of points with angles $0$ and $\pi$. In these neighborhoods we
consider very small outward perturbations of the circle (see 
Figure~\ref{mos.int}). The image of the curve $\Gamma$ is the same as the 
image of the circle, except the neighborhoods of the points with the angles 
$\pi/2$ and $3\pi/2$. Clearly if our perturbation is less than $a$, then the 
image of the curve $\Gamma$ does not intersect with the curve itself. \qed

Notice that in the above example the intersections of the circles centered
at the origin with their images occur at points with angles $0$ and $\pi$,
and they are not transversal. If the intersection is transversal, then there 
exists a neighborhood of the circle such that for any closed curve in this 
neighborhood the intersection property holds. However this neighborhood is 
rather thin and it is simple to change the construction in the proof of 
Lemma~\ref{l:int.prop} to make the intersections to be transversal by adding 
some expansion along the horizontal coordinate with the coefficient $(1+b)$, 
provided $a>b>0$. It is also straightforward to construct the twist map with 
the same property by the superposition of the map $f$ with the rotation by 
the angle $\gamma r$, providing the constant $\gamma$ is small enough.

\bigskip

It is worth noting that the twist property is very essential for the
existence of the invariant curve and one cannot extend above results for
the case of the pure rotation, i.e., when $\Phi(r) = \const$.

\begin{lemma} \label{smooth-inf} Let $\Phi(r) \equiv \alpha = \const$. Then
for any sufficiently small $\ep>0$ there exist a pair of $C^\infty$-smooth
functions $h,g$ satisfying the above assumptions such that the perturbed
system~(\ref{twist-0}) does not have an invariant curve and moreover there
are trajectories going to infinity. \end{lemma}

\proof Consider the following perturbed family of maps, depending on the
parameter $\ep \ge 0$
\beq{twist-2}{
    \function{&\!\!\! \phi \to \phi + \alpha + \ep h \mod1 \\
              &\!\!\! r    \to r    + \ep g(\phi) } }
where $h \in [0,1)$ is a constant and $g(\phi)$ is a $C^\infty$ 1-periodic
function. The system~(\ref{twist-2}) is a skew product of two maps, and the
first (angle) coordinate does not depend on the second (radial) one. Clearly
we can choose the constant $h$ in such a way that the number $\alpha+\ep h$
will be rational. In this case the first map has a periodic trajectory
$\phi_1, \phi_2, \dots , \phi_k$. Now setting $g \equiv 0$ we immediately
see that all the trajectories are periodic (with the same period), and thus 
do not fill densely corresponding invariant circles.

To show that at least some trajectories may go to infinity we need to define
the function $g$ in a bit more delicate way. We define the function $g$ at
the points $\phi_i$ as $+1$ and extend it as a $C^\infty$-function to the
unit circle, such that it rapidly goes to $-1$ in small neighborhoods of the
points $\phi_i$ and is equal to $-1$ otherwise. This definition clearly
implies the intersection property, while the trajectory starting at the point
$(\phi_1, 1)$ goes to infinity with the linear rate $\ep n$. \qed

\section{Twist maps with discontinuous perturbations}

\begin{definition} By an {\em $\ep$-discretization}, $\ep>0$, of a set $X
\subset \IR^d$ we mean a choice of an ordered lattice (a collection of
points) $X_\ep$ in the set $X$ such that the distances between its
neighboring elements do not exceed $\ep$. By an {\em operator of
$\ep$-discretization} \label{i:discr.oper} we mean a map $D_\ep: X \to
X_\ep$ that associates to each point $x \in X$ its nearest point on the
lattice $x_\ep = D_\ep(x) \in X_\ep$ (if there are several such points we
choose the point with the minimal index). Recall that the lattice $X_\ep$
is assumed to be ordered. The value of the parameter $\ep>0$ is called the
{\em diameter} of the $\ep$-discretization or the magnitude of the
corresponding perturbation.  \end{definition} 

\begin{definition} By an {\em $\ep$-discretized} (perturbed) system for the
map $f$ with the phase space $X$ we mean a pair $(f_\ep,~X_\ep)$, where
$f_\ep = D_\ep \circ f$. \end{definition} 

Note that contrary to usual perturbation schemes, the discretized 
(perturbed) system here is given not on the original phase space $X$ 
but on the discrete lattice $X_\ep$. 

In the case of round-off errors in computer modeling we deal with a uniform
$\ep$-discretization, i.e., for $\ep = 1/N$, $N \in \IZ_+$, the lattice
$X_\ep$ consists of all points $x \in X$ with rational coordinates with the
denominator $N$.

Denoting the twist map~(\ref{twist}) by $F$ we may represent the
$\ep$-discretized twist map $D_\ep \circ F$ as follows:
\beq{twist-1}{
    \function{&\!\!\! \phi \to \phi + \Phi(r)
                                    + \frac{\ep}r h(\phi,r,\ep) \mod1 \\
              &\!\!\! r     \to r   + \ep g(\phi,r,\ep) } }
The perturbations due to round-off errors that we consider are
discontinuous, but there is an additional small parameter $1/r$
(for $r\gg1$) in the first relation, which arises due to the fact that the
perturbation of the angle is inversely proportional to the distance from
the origin.

%%%%%%%%%%%%%%%%%%%%%%%%%%%%%%%%%%%%%%%%%%%%%%%%%%%%%%%%%%%%%%%%%%%%%%%%%%%%
%%%%%  Smooth approximations for $D_\ep$
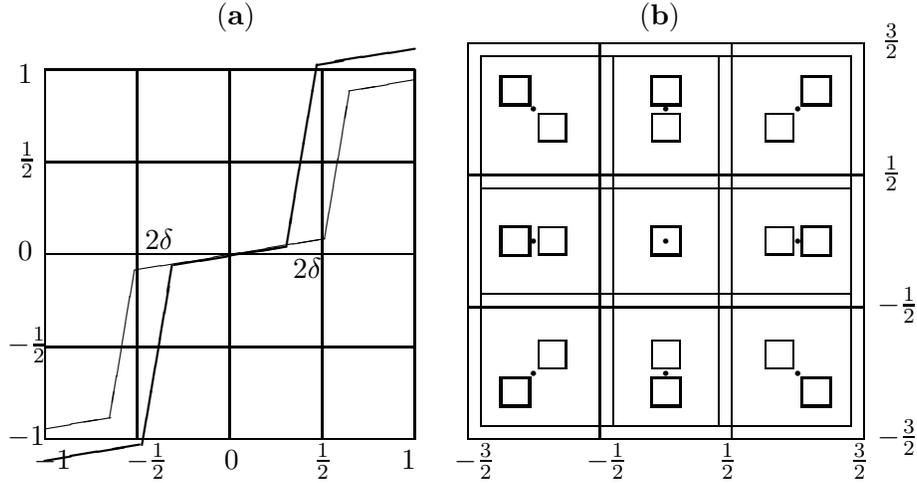
\begin{figure} \begin{center} %\setlength{\unitlength}{1pt}
\begin{picture}(310,150)
   \multiput(5,5)(35,0){5}{\line(0,1){140}}
   \multiput(5,5)(0,35){5}{\line(1,0){140}}
   \put(39,69){\line(-1,-6){9.3}} \put(111,81){\line(1,6){9.3}}
   \put(5,9){\line(6,1){25}} \put(145,141){\line(-6,-1){25}}
   \put(39,69){\line(6,1){72}}
   \thicklines
   \put(5,-3){\line(6,1){36}} \put(54,71){\line(6,1){42.5}}
   \put(145,153){\line(-6,-1){37}}
   \put(42,3){\line(1,6){11.3}} \put(108,147){\line(-1,-6){11.4}}
   \put(1,-6){$-1$} \put(36,-6){$-\frac12$} \put(73,-6){$0$}
   \put(107,-6){$\frac12$} \put(140,-6){$1$}
   \put(-9,3){$-1$} \put(-9,38){$-\frac12$} \put(-5,73){$0$}
   \put(-5,108){$\frac12$} \put(-5,140){$1$} \put(70,162){({\bf a})}
   \put(43,77){$2\delta$} \put(99,66){$2\delta$}
   \put(160,0){\thinlines
            \multiput(5,5)(50,0){4}{\line(0,1){150}}
            \multiput(5,5)(0,50){4}{\line(1,0){150}}
            \multiput(5,5)(0,50){3}
                     {\multiput(0,0)(50,0){3}{\put(25,25){\circle*{2}}}}
            \put(10,10){\line(1,0){140}} \put(10,60){\line(1,0){140}}
            \put(10,100){\line(1,0){140}} \put(10,150){\line(1,0){140}}
            \put(10,10){\line(0,1){140}} \put(60,10){\line(0,1){140}}
            \put(100,10){\line(0,1){140}} \put(150,10){\line(0,1){140}}
            \thicklines
            \multiput(0,0)(0,57){3}
                {\multiput(0,0)(57,0){3}{\put(18,18){\framebox(10,10){}}}}
            \thinlines
            \multiput(0,0)(0,43){3}
                {\multiput(0,0)(43,0){3}{\put(32,32){\framebox(10,10){}}}}
            \put(70,162){({\bf b})}
            \put(0,-6){$-\frac32$} \put(50,-6){$-\frac12$}
            \put(100,-6){$\frac12$} \put(150,-6){$\frac32$}
            \put(160,3){$-\frac32$} \put(160,53){$-\frac12$}
            \put(162,103){$\frac12$} \put(162,153){$\frac32$}
              }
\end{picture} \end{center}
\caption{Approximation for the operator $D_1$. (a) One-dimensional
approximation. (b) Two-dimensional approximation.
By thick (thin) lines we indicate the approximation from above (below).
On figure (b) by thick black dots we indicate integer points, thick-lined
small squares correspond to the images under the approximation from above of
unit rectangles centered at integer points, without $2\delta$-narrow strips
near their boundaries, indicated by thin lines. Thin small squares
correspond to the images under the approximation from below.}
\label{appr-fig} \end{figure}
%%%%%%%%%%%%%%%%%%%%%%%%%%%%%%%%%%%%%%%%%%%%%%%%%%%%%%%%%%%%%%%%%%%%%%%%%%%%%

Our aim is to construct $C^\infty$ smooth approximations $D_\ep^\pm$ for the
discretization operator $D_\ep$. The first approximation $D_\ep^+$ is the
approximation from above in the sense that for any point $x \in \IR^2$
we have
$$ |D_\ep^+(x)| \ge |D_\ep(x)| ,$$
while for the second approximation $D_\ep^-$ -- the approximation from below
the opposite inequality
$$ |D_\ep^-(x)| \le |D_\ep(x)| $$
holds. Both of these approximations should satisfy also the property that
their graphs lie in the $3\delta\ep$ neighborhood of the graph of the
operator $D_\ep$. Here $0<\delta\ll1$ is the parameter of the approximation.
Notice that $D_\ep(x)=\ep D_1(x/\ep)$ for any $\ep>0$ and therefore it is
enough to construct the approximations for $\ep=1$.

We define the the approximation from above $D_1^+$ in three steps. On the
first step we define the following one-dimensional piecewise linear
function, depending on a small parameter $0<\delta\ll1$:
$$ f_1^+(t) := \function{
   \frac{2\delta}{1/2-2\delta}t &\mbox{if } 0 \le t \le 1/2-2\delta \\
   \frac{1-\delta}{\delta} \lrp{t - D_1(t) - \frac12 + 2\delta} + D_1(t)
      + 2\delta &\mbox{if }
       - 2\delta < t-D_1(t)-\frac12 \\
                 & \qquad \le - \delta, \; t>0 \\
   \frac{\delta}{1-\delta} \lrp{t - D_1(t) - \frac12 + \delta} + D_1(t) + 1
      + \delta  &\mbox{if }
       -\frac12 - \delta < t-D_1(t) \\
                & \qquad \le \frac12 - 2\delta, \; t>\frac12 \\
   -f_1^+(-t) &\mbox{if } t<0 }$$
The graph of this function is shown by thick lines on
Figure~\ref{appr-fig}(a). By the construction we have
\bea{ \a f_1^+(t) \le D_1(t) \; \mbox{if } t < 0 \\
      \a f_1^+(t) \ge D_1(t) \; \mbox{if } t \ge 0 .}
Thus $|f_1^+(t)| \ge |D_1(t)|$ for any $t \in \IR^1$. Straightforward
estimates also show that the distance between the graphs of $D_1$ and
$f_1^+$ is equal to $2\delta$.

 From the first sight it seems that the one-dimensional approximation from
above for the discretization operator can be done in a much more simple way,
by the function:
$$ \hat f_1(t) := \function{ 2x &\mbox{if } |x|<1/2 \\
                             x-1/2 &\mbox{if } x<-1/2 \\
                             x+1/2 &\mbox{if } x>1/2 }. $$
However for any twist map with the perturbation $\ep \hat f_1(t/\ep)$ on
each coordinate almost all trajectories go to infinity, and thus there is
no invariant curves.

On the second step we smooth out this piecewise linear function to obtain
a new function having a given number of derivatives. The simplest way to do
it is to apply the following integral smoothing operator:
$$ {\cal S}_\sigma h(t) := \int_{|s|<\sigma} \sigma^{-1}
                      S_\sigma(s) h(t-s) \, ds ,$$
where $0<\sigma\ll1$, and the kernel $S_\sigma \in C^\infty$,
$S_\sigma(t)=0$ for $|t|>\sigma$ and
$$ \int_{-\infty}^\infty t^k S_\sigma(t) \, dt
   = \function{\sigma &\mbox{for } k=0 \\ 0 &\mbox{for } 0<k<l }, $$
where $l$ is a fixed number. It is shown (see details, for example, in
\cite{Mos}) that for any continuous function $h$ the function
${\cal S}_\sigma h \in C^\infty$ and its derivatives up to the order $l$ are
effectively estimated. Moreover we have a good control over the difference
$h-{\cal S}_\sigma h$, which enables to show that if the parameter $\sigma$
is sufficiently small, for example, $\sigma=\delta^2$, then applying this
operator to $f_1^+$ we obtain a function $\tilde f_1^+ \in C^\infty$, being
the approximation of the discretization operator from above. The distance
between the graphs will be somewhat enlarged, but is still less than
$3\delta$.

The multidimensional discretization operator $D_1$ acts on each coordinate
independently of others. Thus on the last step of the approximation
procedure we apply the function $\tilde f_1^+$ on each coordinate
independently, which defines the smooth define approximation $D_1^+$ for the
two-dimensional discretization operator $D_1$. The action of this
two-dimensional transformation, shown on Figure~\ref{appr-fig}(b), is the
following. First, it maps almost linearly unit squares centered at integer
points (except $2\delta$-narrow strips near their boundaries, indicated by
thin lines on the figure) to small squares of size $\delta$, located near
the corresponding integer points and shifted outward with respect to the
origin. Second, the mentioned above strips are expanded to the complement of
these small squares. Since $f_1^+$ is the approximation from above for the
one-dimensional discretization operator, similar property holds also for
$D_1^+$:
$$ |D_1^+(x)| \ge |D_1(x)| $$
for any $x \in \IR^2$. Thus, according to our definition, $D_1^+$ is the
approximation from above for $D_1$. Straightforward estimates also show 
that the distance between the graphs $D_1$ and $D_1^+$ is not more than 
to $3\delta$.

\bigskip

In a similar way one can construct the approximation from below $D_1^-$.
The corresponding one-dimensional version is defined as
$$ f_1^-(t) := \function{
   \frac{\delta}{1/2+\delta}t &\mbox{if } 0 \le t \le 1/2+\delta \\
   \frac{1-3\delta}{\delta} \lrp{t - \frac12 - \delta} + \delta
       &\mbox{if }  \delta \le t - \frac12 < 2\delta \\
   \frac{\delta}{1-\delta} \lrp{t - D_1(t) - \frac12 + 2\delta} + D_1(t)
       - 2\delta  &\mbox{if }
       -\frac12 + 2\delta < t-D_1(t) \\
        & \qquad \le \frac12 + \delta, \; t>\frac12 \\
   \frac{1-\delta}{\delta} \lrp{t - D_1(t) - \frac12 + \delta} + D_1(t)
       - \delta  &\mbox{if }
       \delta < t-D_1(t)+\frac12 \le 2\delta, \\ &\qquad t>\frac12 \\
   -f_1^-(-t) &\mbox{if } t<0 }$$
The graph of this function is shown by thin lines on
Figure~\ref{appr-fig}(a). By the construction we have
\bea{ \a f_1^-(t) \ge D_1(t) \; \mbox{if } t < 0 \\
      \a f_1^-(t) \le D_1(t) \; \mbox{if } t \ge 0 }
for any point $|t|>1$. Thus $|f_1^-(t)| \le |D_1(t)|$ for $|t|>1$, which
makes $f_1^-$ the approximation from below everywhere except the central
interval $[-1,1]$. Notice that there is no continuous one-dimensional
approximation from below with nondegenerate derivatives for $D_1$ for all
$t \in \IR^1$.

The second and the third steps of the procedure are exactly the same as in
the construction of the approximation from above. The scheme of the action
of the approximation from below is shown by thin-lined small squares on
Figure~\ref{appr-fig}(b). It is worth noting that the operator $D_1^-$ does
not satisfy our requirement to be the approximation from below in the
$1+2\delta$-square centered at the origin. However we shall apply our
approximation scheme only for the annulus $A$ with nonzero smaller radius.
Therefore we do not need to define the discretization operator near the
origin. Another problem is to verify the approximation from below property
for unit squares intersecting with the coordinate axes. The operator $D_1^-$
restricted to such squares enlarges one of the coordinates and decreases
another one. However, take an integer point $(0,n)$ and notice that
$$ (n-\delta)^2 + \delta^2 = n^2 - 2\delta(1-\delta) < n^2 .$$
Therefore the emphasized small square lies entirely inside of the circle of
radius $n$ centered at the origin. Thus the effect of the decreasing is
stronger, which proves the desired property.

\bigskip

The proof of the intersection property for the approximation from above is
as follows. Consider a circle centered at the origin with a sufficiently
large radius. Then locally this circle is arbitrary close the the straight
line. Take the segment of the circle with the ``slope'' $-1$. Clearly there
is a point $(\phi,r)$ in the lattice of ``half-integer points''
$\ep(\IZ^2 + (1/2,1/2))$ arbitrary close to this segment in the same
``half-plane'' as the origin. Then by the construction of our approximation
there is an intersection of the image of the circle with the circle itself
near the image of the point $(\phi,r)$. Indeed in the
$2\ep\delta$-neighborhood of the point $(\phi,r)$ the function $D_\ep^+$
takes both values $(\phi_-,r-\ep/2)$ and $(\phi_+,r+\ep/2)$, which
yields the intersection. In the same way we can deal with a curve
sufficiently $C^1$-close to the circle. In the same way one can prove the
intersection property for the approximation from below $D_\ep^-$.

Therefore the both maps $D_\ep^\pm \circ F$ satisfy the conditions of
Theorem~\ref{Moser}, and thus they have invariant curves, defined by the
relations $r=\Gamma_\pm(\phi)$. In spite of the fact that our approximation
is the approximation from above (from below), the angles corresponding to
the points $D_\ep^\pm \circ F(x)$ and $D_\ep \circ F(x)$ are different.
Therefore we cannot apply the idea that every time the trajectory of the
approximating map is further from the origin or at least not closer (on
closer in the case of the approximation from below) than the discretized
one, being at the same time uniformly bounded (by the invariant curves).

To prove the boundedness we need an additional trick. For any point
$x \in \IZ^2$ consider its ``shadow'' point $x^s$ on the invariant curve
(the point with the same angle). Now comparing the images of these points
$D_\ep^\pm \circ F(x^s)$ and $D_\ep \circ F(x)$ we obtain the boundedness
of the trajectory of the discretized map.

This completes the proof of the first part of Theorem~\ref{t:twist}.
\smallskip

To prove the second part we restrict our twist map $F$ to the annulus 
${\cal A}(r_-,r_+)$ such that $1 \ll R \le r_- < 2R \le r_+ \le 3R$ with a 
sufficiently large $R$. Then, after the following change of variables: 
$(\phi,r) \to (\phi, r_- r)$, we obtain the twist map of the new annulus 
${\cal A}(1,r_+/r_-)$ with the perturbation of order $\ep/r_-$, which can 
be made arbitrary small by the choise of the constant $r_-$. Therefore 
the argument in the first part of the proof yields the boundedness of any 
trajectory of the discretized map with arbitrary $\ep<\infty$. On the other 
hand, in a bounded region there is only a finite number of points of the 
discretized lattice, which yields the eventual periodicity. 
Theorem~\ref{t:twist} is proven. \qed

The similar construction can be applied for the ``integer part'' 
discretization, when each point is mapped to its integer part. In spite of 
of some differences, this type of discretizations leads to the same 
eventually periodic behavior. On the other hand, the truncation operator for 
arbitrary small truncation error leads to quite different behavior: almost 
any trajectory of the truncated system goes to zero with time \cite{DKKP}.

\section{Discretization of the twist map in Lobachevski's metric}

In this section we discuss some our findings about the properties of
discretizations of the twist map in Lobachevski's metric. The Lobachevski 
plane $\IL_2$ can be realized as the upper semi-plane $y\ge0$ of the 
Eucledian two-dimensional plane $\IR^2$ with the metric 
$ds^2=((dx)^2+(dy)^2)/y^2$, which we denote by $\rho(\cdot,\cdot)$. 
We shall consider two different discretizations of $\IL_2$: one with respect
to the usual integer lattice $\IZ^2$, and another one with respect to 
$\IZ_L = \{(x,y), x\in\IZ, \; y=2^k, \, k\in\IZ \}$. The reason for the 
introduction of the second lattice is that it is uniform with respect to 
the Lobachevski's metric, which is not the case for the lattice $\IZ^2$.
Notice that the lattice $\IZ_L$ is not locally finite.

Observe that the distance in the Lobachevski's metric between any two 
horizontal straight lines vanishes as $x \to \pm\infty$. Therefore any 
vertical straight line plays a role of the radius from the infinity 
$y=\infty$, while any horizontal straight line is a circle centered at 
the infinity. Consider a map $F:\IL_2 \to \IL_2$ defined as 
$F(x,y) = (x+\Phi(y),y)$. The discretization operator 
$D:\IL_2 \to \IZ^2 \cup \IZ_L$ 
we define as the closest point map in the Lobachevski's metric. 

\begin{prop} The intersection of $\IZ^2$ with any horizontal line with 
the integer coordinate $y \in \IZ$ is invariant with respect to the 
$(\IZ^2 \cup \IZ_L)$-discretized map $D\circ F$. \end{prop}

\proof It suffices to show that $D(x,n)=(m,n)$ for any numbers $x\in\IR$ and 
$n\in\IZ$ and some integer $m\in\IZ$. Let $x=m+\delta$, where $m\in\IZ$ and 
$|\delta|\le1/2$. Then for $D(x,n)=(x',y')$ we have:
$$ \rho((x,n),(x',y')) 
   \ge \left| \int_n^{y'} \frac{dy}{y} \right| = |\ln(y'/n)| .$$
Therefore if $y'\ne n$ and $n>1$, then 
$$ \rho((x,n),(x',y')) \ge |\ln((n+1)/n)| 
   > \frac1n - \frac1{2n^2} > \frac{|\delta|}{n} .$$
On the other hand, $\rho((x,n),(x',y'))<|\delta|/n$ for $(x',y')=(m,n)$.

It remains to check the case $n=1$ and $y'=2^{-k}$ with $k\ge1, k\in\IZ$. 
In this case  
$$ \rho((x,1),(x',y')) \ge \left| \int_{\frac12}^{1} \frac{dy}{y} \right| 
   = \ln2 > |\delta| > \rho((x,1),(m,1)) .$$
\qed

It is of interest that the intersection of the second lattice $\IZ_L$ 
with horizontal lines is not invariant with respect to the discretized 
map. The proof of this statement and the study of the discretization of 
finite radius rotation maps in Lobachevski's metric is technically 
sufficiently complex and will be published elsewhere.

\bigskip \bigskip %%%%%%%%%%%%%%%%%%%%%%%%%%%%%%

\end{document}